\UseRawInputEncoding

\documentclass[12pt]{article}
\usepackage{longtable,supertabular}
\usepackage{amsmath}
\usepackage{amssymb}
\usepackage{latexsym}
\usepackage{cite}

\title{\bf Generalized Singleton Type Upper Bounds}
\author{Hao Chen, Longjiang Qu, Chengju Li, Shanxiang Lyu \\and Liqing Xu
  \thanks{Hao Chen, Shanxiang Lyu and Liqing Xu are with the College of Information Science and Technology/Cyber Security, Jinan University, Guangzhou, Guangdong Province, 510632, China, haochen@jnu.edu.cn, shanxianglyu@gmail.com, lqxu1@jnu.edu.cn. Longjiang Qu is with the College of Liberal Arts and Sciences, National University of Defence Technology, Changsha, Hunan Province 410073, China. Chengju Li is with the Shanghai Key Laboratory of Trustworthy Computing , School of Computer Science and Software Engineering, East China Normal University, Shanghai 200062, China, cjli@sei.ecnu.edu.cn. The research of Hao Chen and Shanxiang Lyu was supported by NSFC Grant 62032009. The research of Shanxiang Lyu was supported by NSFC Grant 61902149. The research of Chengju Li was supported by NSFC Grant 12071138.}}

\begin{document}

\maketitle

\begin{abstract}
In this paper, we give  upper bounds on the sizes of $(d, L)$ list-decodable codes in the Hamming metric space from covering codes with the covering radius smaller than or equal to $d$. When the list size $L$ is $1$, this gives many new Singleton type upper bounds on the sizes of codes with a given minimum Hamming distance. These upper bounds are stronger than the Griesmer bound when the lengths of codes are large. Some upper bounds on the lengths of general small Singleton defect codes or list-decodable codes attaining the generalized Singleton bound are given. As an application of our generalized Singleton type upper bounds on Hamming metric error-correcting codes, the generalized Singleton type upper bounds on insertion-deletion codes are given, which are much stronger than the direct Singleton bound for insertion-deletion codes when the lengths are large. We also give upper bounds on the lengths of small dimension optimal locally recoverable codes and small dimension optimal $(r, \delta)$ locally recoverable codes with any fixed given minimum distance.
\end{abstract}


\section{Introduction}

Let ${\bf F}_q$ be a finite field of the size $q$, where $q$ is a prime power. When $q$ is prime, this finite field is called a prime field. For a vector ${\bf a} \in {\bf F}_q^n$, the Hamming weight of ${\bf a}$ is the number of non-zero coordinate positions. The Hamming distance $d({\bf a}, {\bf b})$ between two vectors ${\bf a}$ and ${\bf b}$ is defined to be the Hamming weight of ${\bf a}-{\bf b}$. The minimum Hamming distance $d({\bf C})$ of a code ${\bf C} \subset {\bf F}_q^n$ is the minimum of Hamming distances $d({\bf a}, {\bf b})$ between any two different codewords ${\bf a}$ and ${\bf b}$ in ${\bf C}$. It is well-known that the minimum Hamming distance of a linear code ${\bf C}$ is the minimum Hamming weight of its non-zero codewords.  Let ${\bf A}_q(n, d)$ be the maximal size of a general code ${\bf C} \subset {\bf F}_q^n$ with the minimum Hamming distance $d$. One of the main problems in the error-correcting coding theory is to determine this ${\bf A}_q(n, d)$. There are many upper bounds on ${\bf A}_q(n, d)$, see Chapter 2 of \cite{HP}. For example it was known that ${\bf A}_q(n, d) \leq q{\bf A}_q(n-1, d)$, see \cite{Quis}. \\

The Singleton bound ${\bf A}_q(n, d)  \leq q^{n-d+1}$ in \cite{Singleton} for a general code with the length $n$ and the given minimum Hamming distance $d$ is the basic upper bound in coding theory, see also \cite{Joshi}. The number $n+1-d-log_q|{\bf C}|$ is called the Singleton defect of a general code ${\bf C} \subset {\bf F}_q^n$. A code attaining this bound or with the zero Singleton defect is called an MDS (maximal distance separable) code.  A code with the Singleton defect $1$ is called almost MDS.  A linear almost MDS code ${\bf C}$, which satisfies the dual code ${\bf C}^{\perp}$ is also almost MDS, is called a near MDS code. \\

To upper bound the maximal lengths of general MDS codes, almost MDS codes or near MDS codes is a classical problem in coding theory and finite geometries, see \cite{Silverman,Alderson,KKO}. The Bush upper bound $${\bf A}_q(q+2, q) \leq q^3-2$$ when $q$ is odd, was given in \cite{Bush} and improved in \cite{Quis}. From this bound it follows that no MDS code of the minimum Hamming distance $q$ over ${\bf F}_q$, $q$ is odd, and of the length $n \geq q+2$  exists. It was proved that the length of an MDS $(n, q^k, n-k+1)_q$ code satisfies $n \leq q+k-1$, see \cite{Silverman,Alderson}. For nontrivial linear MDS codes, the dimension $k$ has to satisfy $k \leq q-1$, then $n \leq 2q-2$, see page 264 of \cite{HP}. The main conjecture of linear MDS codes proposed in \cite{Segre} claims that the length of a linear MDS code over ${\bf F}_q$ is at most $q+1$, except some exceptional cases. In \cite{Ball} the main conjecture was proved for linear MDS codes over prime fields. The main conjecture for linear MDS codes has been proved for small dimensions, small $q$'s and some geometric codes, see \cite{Chen2,ChenYau} and \cite{BTB,Alderson} and references therein. For the recent progress in the main conjecture of linear MDS codes, we refer to \cite{Ball1}. The main conjecture of near MDS codes was proposed in \cite{Landjev}. The maximal lengths of linear almost MDS codes were studied in \cite{Gulati,Boer}. Some upper bounds were given for small $q$'s and the special case such as $d>q$. Some classification results about non-linear MDS codes over small fields were given in \cite{KKO}. We refer to \cite{Ghorpade,Kaipa} for the counting of the number of MDS linear codes. New MDS codes which are not equivalent to Reed-Solomon codes and twisted Reed-Solomon codes were constructed from algebraic curves in \cite{Chen5}. In this paper, our upper bounds on the lengths of MDS codes and small Singleton defect codes are for both linear and non-linear codes.\\

The Griesmer bound for a linear $[n, k, d]_q$ code proposed in \cite{Griesmer} asserts that $$ n \geq \Sigma_{i=0}^{k-1} \lceil \frac{d}{q^i} \rceil,$$ see Section 2.7 of \cite{HP}. This bound is stronger than the Singleton bound for linear codes, since $n \geq d+ \Sigma_{i=1} \lceil \frac{d}{q^i} \rceil \geq d+k-1$  and serves as the best upper bounds for many best known long linear codes in \cite{codetable}.\\

Let ${\bf F}_q$ be an arbitrary finite field, $P_1,\ldots,P_n$ be $n \leq q$ elements in ${\bf F}_q$. The Reed-Solomon codes $RS(n,k)$ is defined by $$RS(n,k)=\{(f(P_1),\ldots,f(P_n)): f \in {\bf F}_q[x],\deg(f) \leq k-1\}.$$ This is a $[n,k,n-k+1]_q$ linear MDS codes based on the fact that a degree $\deg(f) \leq k-1$ polynomial has at most $k-1$ roots. \\

For a code ${\bf C} \subset {\bf F}_q^n$, we define its covering radius  by $$R_{covering}({\bf C})=\max_{{\bf x} \in {\bf F}_q^n} \min_{{\bf c} \in {\bf C}} \{wt({\bf x}-{\bf c})\}.$$ Hence the Hamming balls $B(x, R_{covering}({\bf C}))$ centered at all codewords $x \in {\bf C}$, with the radius $R_{covering}({\bf  C})$ cover the whole space ${\bf F}_q^n$.  We refer to the excellent book \cite{CHLL} on this classical topic of coding theory.  Actually the covering radius $R_{covering}({\bf C})$ of a linear $[n, k]_q$ code ${\bf C} \subset {\bf F}_q^n$ can be determined as follows. If $H$ is any $(n-k) \times n$ parity check matrix of ${\bf C}$, $R_{covering}({\bf C})$ is the least integer such that every vector in ${\bf F}_q^{n-k}$ can be represented as ${\bf F}_q$ linear combinations of $R_{covering}({\bf C})$ or fewer columns of $H$, see \cite{Janwa1,DMP}.  The redundancy upper bound $$R_{covering}({\bf C}) \leq n-k$$ for a linear $[n, k]_q$ code, follows immediately, see \cite{CHLL}. Let $n$ be a fixed positive integer and $q$ be a fixed prime power, for a given positive integer $R<n$, we denote $K_q(n, R)$ the minimal size of a code ${\bf C} \subset {\bf F}_q^n$ with the covering radius smaller than or equal to $R$. Set $ \frac{log_q K_q(n, \rho n)}{n}=k_n(q, \rho)$. The following asymptotic bound is well-known,  $$1-H_q(\rho) \leq k_n(q,\rho) \leq 1-H_q(\rho)+O(\frac{logn}{n}),$$ where $H_q(r)=rlog_q(q-1)-rlog_q r-(1-r)log_q (1-r)$ is the $q$-ary entropy function, see Chapter 12 of \cite[Chapter 12]{CHLL}.  In particular when each vector in ${\bf F}_q^n$ is in exactly one Hamming ball centered in a codeword in ${\bf C}$ with the radius of $R$, we call this code a perfect codes. The existence and determination of perfect codes in various metrics are the fascinating topic in coding theory, related to many other topics of mathematics. Hamming code and Golay code are basic examples of perfect codes, see Chapter 11 of \cite{CHLL}. It is well-known that the covering radius of the Reed-Solomon $[n, k]_q$ code is $n-k$ if $n \leq q$, see \cite{CHLL}. There have been extensive research on covering radii of codes in the Hamming metric. For example, in the paper \cite{DJ1991}, covering radii of more than six thousand binary cyclic codes were calculated and determined. We refer to \cite{EFS} for the recent work on the generalized covering radii of linear codes.\\

Let $({\bf X}, d)$ be a general finite metric space.  A covering code ${\bf C} \subset {\bf X}$ of the radius $R$ means that the balls centered at codewords in ${\bf C}$ with the radius $R$ cover the whole space ${\bf X}$. The covering radius of a code ${\bf C} \subset {\bf X}$ is the smallest radius with this covering property. A code ${\bf C} \subset {\bf X}$ is called (combinatorial) $(d_{list},L)$ list-decodable if  the ball of the radius $d_{list}$ centered at any ${\bf x} \in {\bf X}$ contains at most $L$ codewords of ${\bf C}$. We refer to \cite{Sudan,Johnson} and \cite[Chapter 5]{HP} for the list-decoding algorithm and list-decodable codes in the Hamming metric space ${\bf F}_q^n$.\\

The classical Singleton bound $$|{\bf C}| \leq q^{n-2d_{list}}$$ for the $(d_{list},1)$ list-decodable codes in \cite{Joshi,Singleton} was generalized to $$|{\bf C}| \leq Lq^{n-\lfloor \frac{(L+1)d_{list}}{L}\rfloor}$$ for  $(d_{list}, L)$ list-decodable codes in the paper \cite{ShangguanTamo}. When $L=2$ and $L=3$ the existence of list decodable Reed-Solomon codes attaining this bound was proved in \cite{ShangguanTamo}. An improvement on this bound was recently presented in \cite{GShangT}. The main focus of the bounds in \cite{ShangguanTamo,GShangT} is for list-decodable codes over large fields. The main focus of this paper is for list-decodable codes over small fields.\\

In this paper, we show that any small covering code with the covering radius smaller than or equal to $R$ gives a good upper bound on the sizes of $(R, L)$ list-decodable codes. From classical results on covering radius of various codes, many new upper bounds on the sizes of list-decodable codes are obtained. On the other hand, any small covering code with the covering radius smaller than or equal to $R$ gives a good upper bound on the sizes of codes with the given minimum Hamming distances $\geq 2R+1$. When the lengths are large these upper bounds are stronger than the Griesmer bound for linear codes. Our Singleton upper bounds show that when lengths are larger, the positive integers $t$'s in the Singleton type upper bounds $${\bf A}_q(n, d) \leq q^{n-td}$$ are larger. These upper bounds are suitable to upper bound the sizes of subfield subcodes. From these upper bounds, we give some upper bounds on the lengths of general (both linear and non-linear) MDS codes and small Singleton defect codes. An upper bound on the lengths of list-decodable codes attaining the generalized Singleton bound in \cite{ShangguanTamo} is also presented. We construct some covering codes as dual codes of linear codes with few nonzero weights. As applications, we show that the direct Singleton upper bounds for general insertion-deletion codes are far away from the tightness when the lengths of codes are large. An upper bound on the lengths of small dimension Singleton-optimal locally recoverable codes with any fixed minimum distance is given. We also give an upper bound on the lengths of small dimension optimal $(r, \delta)$ locally recoverable codes with any fixed given minimum distance. This is the first paper which translates classical results about covering codes to upper bound the sizes of general codes of the given minimum distance. \\

\section{Covering code upper bounds}

The following covering code upper bound is general since we can take any code ${\bf C}'$ with the covering radius $R_{covering}({\bf C}) \leq d$. Actually any good upper bound for the covering radius implies a good upper bound on the sizes of $(d_{list}, L)$ list-decodable codes from our covering code upper bound.\\

{\bf Theorem 2.1.} {\em Let ${\bf C} \subset {\bf F}_q^n$ be an  $(d_{list}, L)$ list-decodable code. Suppose that ${\bf C}' \subset {\bf F}_q^n$ is a code with the covering radius $R_{covering} \leq d_{list}$, then we have $$|{\bf C}| \leq L|{\bf C}'|.$$ Hence $$|{\bf C}| \leq L K_q(n,d_{list}).$$}\\

{\bf Proof.} The balls of the radius $d_{list} \geq R_{covering}({\bf C}')$ centered at the codewords of ${\bf C}'$ cover the whole space ${\bf F}_q^n$. Then in each such ball there are at most $L$ codewords of ${\bf C}$. We have $|{\bf C}| \leq L \cdot |{\bf C}'|$. The other conclusions follow directly.\\

The covering upper bounds for a $(d_{list}, L)$ list-decodable code ${\bf C} \subset {\bf F}_q^n$ are weaker than the sphere-packing upper bound$$|{\bf C}| \leq L \cdot \frac{q^n}{\Sigma_{j=0}^{d_{list}} \displaystyle{n \choose j}(q-1)^j}.$$ This sphere-packing upper bound was well-known. However when $n$ and $d_{list}$ are large, it is computationally infeasible to get clear and explicit upper bounds from this expression. Hence by using various results in covering codes, explicit upper bounds for list-decodable codes in the Hamming metric space can be obtained. In this section some such upper bounds on length $n$ list-decodable codes are obtained for concrete length $n$.\\

From Theorem 2.1 and the asymptotic bound of covering codes, it follows immediately that for a sequence of $[n_i, k_i]_q$  codes ${\bf C}_i \subset {\bf F}_q^{n_i}$ which is $(rn_i, L_i)$ list-decodable, where $n_i$ goes to the infinity, and the rate $\lim \frac{k_i}{n_i} \geq 1-H_q(r)+\epsilon$, then the list sizes $L_i$ are exponentials of $n_i$. Then this sequence of codes are non list-decodable.\\

There are a lot of classical results about the binary covering codes, we refer to \cite{CHLL,Litsyn,DJ1991,BDMP}. From the table \cite{Litsyn} of Litsyn, we have the following upper bound on the length $16$ binary $(3, L)$ list decodable code ${\bf C} \subset {\bf F}_2^{16}$, $$|{\bf C}| \leq 192L,$$ since $K_2(16,3) \leq 192$.  Thus if the linear $[16, 9, 4]_2$ code is $(3, L)$ list-decodable, the list size has to be at least $3$.\\

The almost trivial redundancy upper bound $R_{covering} \leq n-k$ for the covering radius of the linear $[n,k]_q$  code, see page 217 of \cite{CHLL}, and our covering code upper bounds imply an upper bound which is close to the generalized Singleton upper bound in \cite{ShangguanTamo}, when the list size is large.\\

{\bf Corollary 2.1 .} {\em  An $(d_{list},L)$ list-decodable code ${\bf C} \subset {\bf F}_q^n$ have to satisfy $$|{\bf C}| \leq Lq^{n-d_{list}}.$$}\\

{\bf Proof.} Let ${\bf C}'$ be a linear $[n, k]_q$ code in ${\bf F}_q^n$, then $R_{covering}({\bf C}') \leq n-k$. This follows from the fact that for any $k$ information set coordinate positions $1 \leq i_1<i_2 \cdots <i_k\leq n$, the coordinates at these positions of codewords in ${\bf C}'$ can be arbitrary vectors in ${\bf F}_q^k$.  For the second conclusion, we take a linear $[n, n-d_{list}]_q$ code ${\bf C}'$, then $R_{covering}({\bf C}') \leq d_{list}$. The conclusion follows immediately.\\

This is weaker than the generalized Singleton upper bound $$|{\bf C}| \leq Lq^{n-\lfloor \frac{(L+1)d_{list}}{L}\rfloor}.$$  in \cite{ShangguanTamo}. However when the list size $L \geq d_{list}$, this almost trivial upper bound is equivalent to the generalized Singleton upper bound in \cite{ShangguanTamo}. \\

From the known linear perfect codes such as linear $[\frac{q^m-1}{q-1}, \frac{q^m-1}{q-1}-m, 3]_q$ Hamming perfect code, binary Golay $[23,12,7]_2$ perfect code and ternary Golay $[11, 6, 5]_3$ perfect code, see, Chapter 11, \cite{CHLL}, we have the following result.\\

{\bf Corollary 2.2.} {\em 1) Let $n=\frac{q^m-1}{q-1}$, where $q$ is a fixed prime power and $m=2,3,4, \ldots$. Let $k >n-m$ be a positive integer. Then if a linear $[n,k]_q$ code  is $(1, L)$ list-decodable, then the list size has to satisfy $L \geq q^{k-n+m}$.\\
2) If a binary linear $[23,k]_2$ code is $(3, L)$ list-decodable, then the list size has to satisfy $L \geq 2^{k-12}$.\\
3) If a ternary linear $[11, k]_3$ code is $(2, L)$ list-decodable, then the list size has so satisfy $L \geq 3^{k-6}$.}\\

It is well-known that the covering radius of the first order $[2^m, m+1, 2^{m-1}]_2$ Reed-Muller code is $2^{m-1}-2^{\frac{m-2}{2}}$ when $m$ is even, see page 243 of \cite{CHLL}. Then the following upper bound for binary $(rn, L)$ list-decodable codes, where $\frac{1}{2}-\frac{1}{2^{\frac{m+2}{2}}} \leq r <\frac{1}{2}$, follows from our covering code bound. The second conclusion follows from Theorem 11.5.3 in page 450 of \cite{HP}. The third conclusion follows from covering radius upper bound for dual BCH codes in \cite{Bazzi}.\\

{\bf Corollary 2.3.} {\em 1) Let $n=2^m$ and $m=2,4,6, \ldots$. Let $r$ be a positive real number satisfying $\frac{1}{2}-\frac{1}{2^{\frac{m+2}{2}}} \leq r <\frac{1}{2}$. If a  binary length $n=2^m$ code ${\bf C}$ is $(rn, L)$ list-decodable, then $$|{\bf C}| \leq L 2^{m+1}. $$ Hence a length $2^m$ binary linear code with the minimum Hamming weight $2^m-2^{\frac{m}{2}}+1$ has its dimension at most $m+1$. \\
2) Moreover for any binary $(s-1, L)$ list-decodable code ${\bf C}$ in ${\bf F}_2^{2s+7}$ where $s$ is any positive integer, we have $|{\bf C}| \leq 64L$. \\
3) Let the length be $n=2^m-1$,  then the cardinality of an $((\frac{1}{2}-(1-o(1))\sqrt{\frac{slog_2n}{n}})n, L)$ list-decodable code ${\bf C} \subset {\bf F}_2^n$, where $s\leq 2^{\frac{m}{2}-1}$, satisfies $$|{\bf C}| \leq L 2^{sm}.$$}\\

For the positive list size $L\geq 2$ we have $\frac{L+1}{L} \leq \frac{3}{2}$, hence the generalized Singleton upper bound in \cite{ShangguanTamo} for this case is at least $$L 2^{2^m -3\cdot 2^{m-2}}=L 2^{2^{m-2}}.$$ Our upper bound is an exponential $2^{\frac{n}{4}-log n+1}$  improvement of the bound in \cite{ShangguanTamo} for binary codes.\\

Let $r$ and $R$ be two fixed positive integers satisfying $r\geq R$. Let $l_q(r,R)$ be the smallest length of a linear $q$-ary code with the covering radius $R$ and the redundancy $r=n-k$.  It is clear that there are  $[n, n-r]_q$  code with the covering radius $R$ for each integer $n \geq l_q(r,R)$. Actually this can be proved directly from the the characterization of  the covering radius of linear codes from its parity check matrix. There are a lot of results about this topic, we refer to \cite{DMP1,DMP,BDMP} and references therein.  From our covering code upper bounds the size of an $(R, L)$ list-decodable code ${\bf C}$ of length $n$ satisfying $n \geq l_q(r,R)$ has to satisfy $$|{\bf C}| \leq L q^{n-r}.$$ When the list size is not one, and $r \geq \frac{3}{2}R$, the upper bound is better than the generalized Singleton bound in \cite{ShangguanTamo} . For example for $r=5>\frac{3}{2} \cdot 3$, the size of a length $n \geq l_q(5,3)$ $(3,L)$ list-decodable code satisfies $|{\bf C}| \leq L q^{n-5}$.  In this case $l_q(5,3) < 2.884q^{\frac{2}{3}} (ln q)^{\frac{1}{3}}$ was proved in \cite{BDMP}. More generally the following result follows from the bound in \cite{DMP}.\\

{\bf Corollary 2.4.} {\em Let $q \geq 8$ be an even prime power. Let $R \geq 4$ be a positive integer. Set $m=\lceil log_q(R+1) \rceil +1$. For any given positive integer $t \geq 3m+2 $, and the length $n \geq Rq^{(t-1)R}+2q^{t-2}+\Sigma_{j=3}^{m+2} q^{t-j}$, the size of an $(R, L)$ list-decodable code ${\bf C} \subset {\bf F}_q^n$ has to satisfy $$|{\bf C}| \leq L q^{n-tR}.$$}

{\bf Proof.} The conclusion follows from Theorem 8 in \cite{DMP}.\\

We observe that when $t \geq 3$, the $(R, L)$ list-decodable codes in Corollary 2.4 can not attain the generalized Singleton bound in \cite{ShangguanTamo}, since $\frac{(L+1)R}{L} < 3R$. Hence this gives an upper bound on the lengths of codes attaining the generalized Singleton bound in \cite{ShangguanTamo}.\\

\section{Covering codes from linear codes with few nonzero weights and the related upper bounds}

One existing point of the classical theory of covering codes is the upper bounds on the covering radius of a code from its dual code. This originated from the work of \cite{Delsarte}. We refer to \cite{Delsarte,Janwa} for some upper bounds on the covering radii from  dual codes. The Delsarte upper bound on the covering radius of a linear code ${\bf C}$ asserts $$R_{covering}({\bf C}) \leq s,$$ where $s$ is the number of nonzero weights of its dual code, see \cite{HP} page 440 Theorem 11.3.3. Hence we have the following result.\\

{\bf Corollary 3.1 (Delsarte type bound).} {\em  Let ${\bf C} \subset {\bf F}_q^n$ be an  $(d_{list}, L)$ list-decodable code. Suppose that ${\bf C}_1 \subset {\bf F}_q^n$ is a linear code such that the total number of nonzero weights of ${\bf C}_1^{\perp}$ is smaller than or equal to $d_{list}$, then we have $$|{\bf C}| \leq L|{\bf C}_1|.$$}\\

The construction and the combinatorial structure of linear codes with few nonzero weights are the classical topic in coding theory, Boolean functions and sequences, we refer to \cite{Kasami,CK,CCZ,Ding1,LLQ,DingDing,Ding2,Ding,MS20,WZD1,LLHQ,XLZD} for many such linear code constructions and the study of their combinatorial structures. The binary simplex code and the Kasami code are typical such linear codes. From the Delsarte upper bound on the covering radius, we have a lot of good covering codes over small fields as in the following examples. The length function $l_q(r, R)$ for covering codes has been studied in \cite{Davy1,Davy,DMP, DMP1}. Actually these construction of linear codes with few nonzero weights can be thought as upper bounds on this length function $l_q(r, R)$ for covering codes.\\

{\bf Example 3.1.} The Kasami codes proposed in  \cite{Kasami} are binary linear $[2^{2m}-1, 3m, 2^{2m-1}-2^{m-1}]_2$ codes with three nonzero weights, $w_1=2^{2m-1}-2^{m-1}$, $w_2=2^{2m-1}$ and $w_3=2^{2m-1}+2^{m-1}$, $m=1, 2, \ldots,...$. Its dual is a binary linear $[2^{2m}-1, 2^{2m}-1-3m]_2$ code with the covering radius at most $3$.  Therefore $l_2(3m, 3) \leq 2^{2m}-1$. These dual codes of Kasami codes as covering codes can be compared with linear covering codes with the covering radius $3$ in \cite{Davy1,Davy}. \\

{\bf Example 3.2.} Some binary linear $[2^{m-1}-1, m]_2$ codes with three nonzero weights were constructed in \cite{DingDing}, $m=2, 3, \ldots$ Their dual codes are linear codes with the covering radius at most $3$ and the codimension $m$. Then $l_2(m, 3) \leq 2^{m-1}-1$.\\

Let $m$ be an odd positive integer. Then a linear $[2^{m-1}, m+1]_2$ code with $4$ nonzero weights was constructed in \cite{Ding2}. The dual code is a binary linear $[2^{m-1}, 2^{m-1}-m-1]_2$ code with the covering radius at most $4$. Then $l_2(m+1, 4) \leq 2^{m-1}$.\\

{\bf Example 3.3.} Let $m$ be an even positive integer and $k|m$ be a positive divisor of $m$ satisfying $k \neq m$ and $k \neq \frac{m}{2}$. A binary linear $[2^{m-1}, m+k]_2$ code with $3$ nonzero weights was constructed in Theorem 21, page 5143 of \cite{WZD1}. The dual code is a binary linear $[2^{m-1}, 2^{m-1}-m-k]_2$ code with the covering radius at most $3$. Then $l_2(m+k, 3) \leq 2^{m-1}$. This is weaker than the bound in Example 3.2.\\

Let $m$ be an even positive integer of the form $2(2k+1)$ and $t$ be a positive divisor satisfying $t \neq 2^{m/2}+1$, $lcm(t, 3)|2^{m/2}+1$, $3|t$. Then a binary linear $[\frac{2^m-1}{t}, m+1]_2$ code with $5$ nonzero weights was constructed in Theorem 24, page 5144 of \cite{WZD1}. The dual code is a binary linear $[\frac{2^m-1}{t}, \frac{2^m-1}{t}-m-1]_2$ code with the covering radius at most $5$. Then $l_2(m+1, 5) \leq \frac{2^m-1}{t}$.\\

Let $m$ be an even positive integer and $t$ be a positive divisor satisfying $t \neq 2^{m/2}+1$, $lcm(t, 3)|2^{m/2}+1$, and $t$ is not a multiple of $3$. Then a binary linear $[\frac{2^m-1}{t}, m+2]_2$ code with $6$ nonzero weights was constructed in Theorem 24, page 5144 of \cite{WZD1}. The dual code is a binary linear $[\frac{2^m-1}{t}, \frac{2^m-1}{t}-m-2]_2$ code with the covering radius at most $6$. Then $l_2(m+2, 6) \leq \frac{2^m-1}{t}$.\\

{\bf Example 3.4.} Let $m$ be a positive integer satisfying $m \equiv 0$ $mod$ $6$. Then a binary cyclic $[2^m-1, \frac{5m}{2}]_2$ code with $7$ nonzero weights was constructed in \cite{WZD}. From the Delsarte upper bound, the dual code is a binary linear $[2^m-1, 2^m-1-\frac{5m}{2}]_2$ code with the covering radius at most $7$. Then $l_2(\frac{5m}{2}, 7) \leq 2^m-1$ for $m=6, 12, \ldots$.\\

{\bf Example 3.5.} Let $q$ be an odd prime power and $m$ be an odd positive integer. Then a linear $[q^m-1, 2m]_q$ code with $3$ nonzero weights was constructed in Theorem 1 of \cite{LLQ}. The dual code is a linear $[q^m-1, q^m-1-2m]_q$ code with the covering radius at most $3$. Then $l_q(2m, 3) \leq q^m-1$.\\

Let $q$ be an odd prime power and $m$ be an even positive integer. Then a linear $[q^m-1, 2m]_q$ code with $4$ nonzero weights was constructed in Theorem 1 of \cite{LLQ}. The dual code is a linear $[q^m-1, q^m-1-2m]_q$ code with the covering radius at most $4$. Then $l_q(2m, 4) \leq q^m-1$.\\

Let $q$ be an odd prime power and $m$ be an odd positive integer. Then a linear $[q^m, 2m+1]_q$ code with $4$ nonzero weights was constructed in Theorem 2 of \cite{LLQ}. The dual code is a linear $[q^m, q^m-1-2m]_q$ code with the covering radius at most $4$. Then $l_q(2m+1, 4) \leq q^m$.\\

Let $q$ be an odd prime power and $m$ be an even positive integer. Then a linear $[q^m, 2m+1]_q$ code with $6$ nonzero weights was constructed in Theorem 2 of \cite{LLQ}. The dual code is a linear $[q^m, q^m-1-2m]_q$ code with the covering radius at most $6$. Then $l_q(2m+1, 6) \leq q^m$.\\

Let $q=p^m$ be an odd prime power with the even exponent $m$, then a linear $[\frac{q-1}{2}, m, \frac{(p-1)(q-\sqrt{q})}{2p}]_p$ code with two nonzero weights is explicitly given in Corollary 4 of \cite{Ding}. Hence for any $(2,L)$ list-decodable code ${\bf C} \subset{\bf F}_p^{\frac{q-1}{2}}$, we have $$|{\bf C}| \leq L p^{\frac{q-1}{2}-m},$$ from the above Delsarte type  upper bound. \\

From Theorem 1 of \cite{XLZD} we get the following upper bound from the above Delsarte type upper bound on length $q^2-1$ code over ${\bf F}_p$, $q=p^m$. $h$ is a fixed positive integer satisfying $h \neq 0$ $mod$ $q+1$, $e=\gcd(h,q+1)$, $t \leq \frac{q+1}{2e}$. Then when $p=2$ the cardinality of an $(2t+1, L)$ list-decodable code ${\bf C} \subset {\bf F}_2^{2^{2m}-1}$ has to satisfy $$|{\bf C}| \leq L \cdot 2^{2^{2m}-1-(2t+1)m}.$$ When $p$ is an odd prime, then the cardinality of an $(2t, L)$ list-decodable code ${\bf C} \subset {\bf F}_p^{p^{2m}-1}$ has to satisfies $$|{\bf C}| \leq L \cdot p^{p^{2m}-1-2tm}.$$

\section{Generalized Singleton type upper bounds}

These upper bounds obtained in Section 2 and 3 for list-decodable codes can be applied to $(\frac{d-1}{2}, 1)$ list-decodable codes.  We can obtain many upper bounds on the sizes of codes with the given minimum Hamming distance $d$. It is verified in this section that these upper bounds are much better than the Griesmer bounds when lengths are larger. By combining the best known covering codes in \cite{Litsyn}, we show that our generalized Singleton type upper bounds are strong for presently best known codes in \cite{codetable}. For codes of the given minimum Hamming distance over a small field, the classical Singleton bound is far away from the tightness. In many cases, the Griesmer bound was used to get the upper bound on its sizes, see the table \cite{codetable} of presently known best codes. \\

{\bf Generalized Singleton type upper bound I.} {\em If there is a covering code ${\bf C}' \subset {\bf F}_q^n$ with the covering radius $R$. Then the size of a code ${\bf C} \subset {\bf F}_q^n$ with the minimum Hamming distance $2R+1$ satisfies $$|{\bf C}| \leq |{\bf C}'|.$$}\\

We can compare the above upper bound for small length best known binary codes in \cite{codetable}. For example from Litsyn's table \cite{Litsyn}, $K_2(15,3)=112$. Then from the general Singleton bound I $M \leq 112$ for an $(15, M, 7)_2$ code. The presently best known code is a linear $[15, 5, 7]_2$ code, see \cite{codetable}. The ratio is $\frac{K_2(15,3)}{2^5}\leq 4$. From Litsyn's table \cite{Litsyn}, for a binary $(33, M, 9)_2$ code, $M \leq K_2(33, 4) \leq 2^{17} \cdot 3$ from the generalized Singleton bound I. The presently best known code is a linear $[33, 14, 9]_2$ code. The ratio is $\frac{K_2(33, 4)}{2^{14}}=24$. Similarly for a binary $(33, M, 11)_2$ code, $M \leq 2^{13} \cdot 11$ from the best known covering code in \cite{Litsyn}. The presently best known binary code is a linear $[33, 12, 11]_2$ code. The ratio is $\frac{K_2(33, 5)}{2^{12}}=22$. From the generalized Singleton bound I, for a $(32, M, 7)_2$ code $M \leq 2^{20}$. The presently best known code is a binrary $[33, 17, 8]_2$ code, see \cite{codetable}. The ratio is $\frac{K_2(32, 7)}{2^{17}}=8$. This illustrates that the above generalized Singleton bound I is strong even for small length binary codes, since the ratio $\frac{K_2(n, R)}{2^k}$ is small where $k$ is the dimension of the best known $[n, k, 2R+1]_2$ code.\\

The following result follows from the binary linear codes with few nonzero weights constructed in \cite{XLZD}.\\

{\bf Corollary 4.1.} {\em Let $R$ be an odd positive integer satisfying $R \leq 2(2^m+1)+1$. An $(n, M, 2R+1)_2$ code of the length $n \geq 2^{2m}-1$ has to satisfy $$M \leq 2^{n-Rm}.$$}\\

Actually any good upper bound on the length function $l_q(r, R)$ studied in \cite{Davy,DMP,DMP1} gives a generalized Singleton type upper bound as follows.\\

{\bf Generalized Singleton type upper bound II.} {\em Let $l_q(r, R)$ be the length function for covering codes. Then an $(n, M, 2R+1)_q$ code of the length $n\geq l_q(r, R)$ has to satisfy $M \leq q^{n-r}$.}\\

{\bf Proof.} Let ${\bf C}$ be a linear covering code of the length $l_q(r, R)$, covering radius $R$ and the codimension $r$. For any nonnegative integer $s \geq 0$, the linear  covering code ${\bf C} \times {\bf F}_q^s$ of length $l_q(r, R)+s$ has its covering radius $R$ and codimension $r$. The conclusion follows immediately.\\

There are some upper bounds on the length function $l_q(tR, R)$ and $l_q(tR+1, R)$, see \cite{DMP,DMP1}. Then if $t\geq 4+\epsilon$, where $\epsilon>0$ is a small positive real number, the generalized Singleton bound for codes of the length at least $l_q(tR, R)$ is stronger than the Griesmer bound. The following corollary follows from the upper bound on the length function for covering codes in \cite{DMP1}.\\

{\bf Corollary 4.2.} {\em Let $q$ be an arbitrary prime power. Let $t$ and $R$ be two positive integers satisfying $R \geq 3$ and $t \geq 1$. Then an $(n, M, 2R+1)_q$ code of the length $n \geq cq^{\frac{(t-1)R+1}{R}} \cdot (ln q)^{\frac{1}{R}}$ has to satisfy $$M \leq q^{n-tR-1}.$$ Here $c$ is a constant independent of $q$.}\\

{\bf Generalized Singleton type upper bound III.} {\em Let $p$ be an odd prime and $m$ be a positive integer. Let $R$ be an even positive integer satisfying $R \leq 2(p^m+1)$. An $(n, M, 2R+1)_p$ code of the length $n \geq p^{2m}-1$ has to satisfy $$M \leq p^{n-Rm}.$$}\\

This upper bound follows from the linear codes with $R$ nonzero weights constructed in \cite{XLZD}.\\

Let $q$ be a prime power and $m$ be a positive integer. Let $e\geq 2$ be a divisor of $q^m-1$ and $a$ be a positive integer satisfying $(q^m-1)$ is not a divisor of $a$. Let $\Delta_1, \ldots, \Delta_e$ be $e$ nonnegative integers satisfying $\Delta_i \neq \Delta_j$ $mod$ $e$, for $i \neq j$, and $\gcd(\Delta_2-\Delta_1, \ldots, \Delta_e-\Delta_1, e)=1$. For example we can take  $\Delta_i=i$. Set $a_i=a+\frac{q^m-1}{e} \Delta_i$, $\delta=\gcd(q^m-1, a_1, \ldots, a_e)$, and $n_1=\frac{q^m-1}{\delta}$. Let $N=\gcd(\frac{q^m-1}{q-1}, ae)$. We assume the condition $N=1$ is satisfied. Then a linear cyclic $[n_1, em]_q$ code with $e$ nonzero weights was constructed in \cite{YXDL} Theorem 7.\\

{\bf Generalized Singleton type upper bound IV.} {\em An $(n, M, 2e+1)_q$ code of the length $n \geq n_1$ has to satisfy $$M \leq q^{n-em}.$$}\\

We can compare this bound with the Griesmer bound for a linear $[n, k, d]_q$ code. The Griesmer bound claims $$n \geq \Sigma_{i=0}^{k-1} \lceil \frac{d}{q^i} \rceil.$$ From the Griesmer bound the right side is smaller than or equal to $2d+k-1$. That is, we can get at most $k \leq n-2d+1$ from the Griesmer bound for a linear $[n, k, d]_q$ code. However when $t \geq 5$, our above generalized Singleton type bounds is much stronger than the Griesmer bound when the length is large. The above generalized Singleton type bounds are suitable to upper bound the sizes of various subfield subcodes.\\

{\bf Example 4.1.} We consider the subfield subcode ${\bf C}|_{{\bf F}_2}$ of Reed-Solomon $[2^{2m}, 2k, 2^{2m}-2k+1]_{2^{2m}}$ code ${\bf C} \subset {\bf F}_{2^{2m}}^{2^{2m}}$, where $k$ is a positive integer satisfying $k \geq 2^{2m-1}-2^m$. This is a binary linear $[2^{2m}, k_1, \geq 2^{2m}-2k+1]_2$ code with the unknown dimension $k_1$. It follows from Corollary 4.1, $$k_1 \leq 2^{2m}- m(2^{2m-1}-k).$$\\

{\bf Generalized Singleton type upper bound V.} {\em Let $m$  and $u$ be two positive integers satisfying $m \geq 3$. Then the cardinality of an $(n, M, 2u+1)_q$ code of the length $n \geq u \cdot \frac{q^m-1}{q-1}$ satisfies $M \leq q^{n-mu}$.}\\

{\bf Proof.} Let ${\bf H}_m$ be an $m \times \frac{q^m-1}{q-1}$ matrix over ${\bf F}_q$, with $\frac{q^m-1}{q-1}$ columns taking all nonzero vectors in ${\bf f}_q^m$. Let ${\bf H}$ be a $um \times \frac{u(q^m-1)}{q-1}$ matrix of the following form.\\

$$
\left(
\begin{array}{ccccccccccc}
{\bf H}_m&{\bf 0}&{\bf 0}&\cdots&\cdots&{\bf 0}\\
{\bf 0}&{\bf H}_m&{\bf 0}&\cdots&\cdots&{\bf 0}\\
{\bf 0}&{\bf 0}&{\bf H}_m&\cdots&\cdots&{\bf 0}\\
\cdots&\cdots&\cdots&\cdots&\cdots&\cdots\\
{\bf 0}&{\bf 0}&{\bf 0}&\cdots&\cdots&{\bf H}_m\\
\end{array}
\right)
$$

Then ${\bf H}$ is a full-rank matrix such that each vector in ${\bf F}_q^{um}$ can be represented as the sum of at most $u$ columns of ${\bf H}$. Let ${\bf C}$ be the linear $[\frac{u(q^m-1)}{q-1}, \frac{u(q^m-1)}{q-1}-um]_q$ code defined by the parity-check matrix ${\bf H}$. Then the covering radius of ${\bf C}$ is at most $u$. On the other hand if $n=\frac{u(q^m-1)}{q-1}+n_1$ satisfying $n_1 <\frac{q^m-1}{q-1}$, we take the covering code ${\bf C} \times {\bf F}_q^{n_1} \subset {\bf F}_q^n$ with the covering radius $u$. The conclusion follows from Theorem 2.1 immediately.\\

Actually from examples in the previous section, we can list more generalized Singleton upper bounds on the sizes of codes with a given minimum Hamming distance.\\

\section{Lengths of general MDS codes and small Singleton defect codes}

The construction of MDS codes is a fundamental problem. For example, there have been a lot of efforts in the  construction of MDS quantum codes, MDS entanglement-assisted quantum codes, the construction of MDS codes in symbol-pair metric and the construction of Singleton-optimal locally recoverable codes, see \cite{Chen4,KZL15,CFXHF,CMST} and references therein. Thus it is always interesting to give an upper bound on lengths of various MDS codes attaining various Singleton bounds. In this section, we give some upper bounds on the length of general MDS codes and almost MDS codes (then near MDS codes) in the Hamming metric. Actually our method is not restricted to general MDS or almost MDS codes satisfying that $n+1-(k+d)=1$, our upper bounds work for general small Singleton defect codes.\\

{\bf Theorem 5.1.} {\em If there exists a linear $[n, n-d]_q$ code with the covering radius $\lfloor \frac{d-1}{2} \rfloor$, then the length of a general MDS code can not be bigger than $n$.}\\

{\bf Proof.} It is clear that we have a linear $[n+s, n-d+s]_q$ covering code ${\bf C} \times {\bf F}_q^s \subset {\bf F}_q^{n+s}$, $ s\geq 0$,  with the covering radius $\lfloor \frac{d-1}{2} \rfloor$. From the generalized Singleton bound I, an $(n+s, M, d)_q$ code has to satisfy $$M \leq q^{n+s-d}.$$ This is not an MDS code.\\

From Theorem 5.1 if there could exist a covering $[q+2, q+2-d]_q$ code ${\bf C} \subset {\bf F}_q^{q+2}$ with the covering radius $\lfloor \frac{d-1}{2} \rfloor$ for $d=1, 2, \ldots, \frac{q+2}{2}$, then the main conjecture follows immediately. However this claim seems too strong for covering codes.\\

From the generalized Singleton bound III, an $(n, M, 2R+1)_q$ code with the length $n \geq l_q(2R+1, R)$ has to satisfy $M \leq q^{n-2R-1}$, that is, the code is not MDS. Hence an MDS $(n, q^{n-d+1}, d)_q$ code, $d=2R+1,2R+2$, can not have length $n \geq l_q(2R+1, R)$. From the bound in Corollary 4.2, an $(n, M, 2R+1)_q$ MDS code or an $(n, M, 2R+2)_q$ MDS code can not have the length $n \geq c q^{\frac{R+1}{R}} \cdot (ln q)^{\frac{1}{R}}$, where $c$ is a constant independent of $q$. In these upper bounds on lengths of general MDS codes, the upper bounds depend on the minimum Hamming distances.\\

From generalized Singleton bounds III and IV in the case $m \geq 3$, we can have upper bounds on the lengths of an $(n, M, d)_q$ MDS code with the minimum distance $d \leq 2(p^m+1)$ or $\frac{d-1}{2}$ is a divisor of $q^m-1$.\\

{\bf Theorem 5.2.} {\em Let $q$ be a prime power, $m$ be a positive integer satisfying $m \geq 3$ and $e \geq 2$ be a divisor of $q^m-1$. We assume that the condition $N=1$ in Section 5 is satisfied. Then an $(n, M, 2e+1)_q$ MDS code or an $(n, M, 2e+2)_q$ MDS code can not have the length $n \geq \frac{q^m-1}{\delta}.$}\\

{\bf Corollary 5.1.} {\em Let $e \geq 2$ be an odd divisor of $q-1$ satisfying $\gcd(e, 3)=1$. Then an $(n, M, 2e+1)_q$ MDS code or almost MDS code can not have length $n \geq q^3-1$.}\\

{\bf Proof.} Set $m=2$. We have $\gcd(q^2+q+1, q-1) \leq 3$. Then $\gcd(\frac{q^3-1}{q-1}, ae)=1$ if we set $a=1$. The conclusion follows from Theorem 5.2 immediately.\\

In \cite{Boer}, it was proved that for a linear $[n, k, d]_q$ code with the Singleton defect $n+1-k-d=s$, and satisfying $d>q$ the length $n$ has to satisfy $$n \leq d-2+\frac{2(q^{s+1}-1)}{q-1}.$$ Corollary 5.1 can be thought as an upper bound on the lengths of MDS codes and almost MDS codes with small Hamming distances.  Moreover our bounds are for both linear and non-linear general MDS codes and almost MDS codes.\\

When the list size is $2$ or $3$, Reed-Solomon codes with $n<<q$ attaining the the generalized Singleton bound $|{\bf C}| \leq L q^{n-\lfloor \frac{(L+1)R}{L}\rfloor}$ were constructed in \cite{ShangguanTamo}. Since for any list size $L$, $\frac{(L+1)R}{L} <2R+1$, we have the following upper bound on the lengths of $(R, L)$ list-decodable codes attaining the generalized bound in \cite{ShangguanTamo}, from Corollary 4.2.\\

{\bf Theorem 5.3.} {\em Let $q$ be a prime power and $R$ be a fixed positive integer satisfying $R \geq 3$. If ${\bf C}$ is an $(R, L)$ list-decodable code attaining the generalized Singleton bound $|{\bf C}|=L \cdot q^{n-\lfloor \frac{(L+1)R}{L}\rfloor}$. Then the length $n$ has to satisfy $$n \leq l_q(2R+1, R) \leq c q^\frac{R+1}{R} ln q^{\frac{1}{R}}.$$}\\

\section{Generalized Singleton type upper bounds on insertion-deletion codes}

The insdel distance $d_{insdel}({\bf a}, {\bf b})$ between two vectors ${\bf a}$ and ${\bf b}$ in ${\bf F}_q^n$ is the minimal number of insertions and deletions which are needed to transform ${\bf a}$ into ${\bf b}$. Actually it was well-known, see  \cite{Cheng2021}, that $$d_{insdel}({\bf a}, {\bf b})=2(n-l),$$ where $l$ is the length of a longest common subsequence of ${\bf a}$ and ${\bf b}$. This insdel distance $d_{insdel}$ is indeed a metric on ${\bf F}_q^n$. It is clear $d_{insdel}({\bf a}, {\bf b}) \leq 2d({\bf a}, {\bf b})$ since  $l \geq n-d({\bf a}, {\bf b})$ is valid for arbitrary two different vectors ${\bf a}$ and ${\bf b}$ in ${\bf F}_q^n$. The insdel distance $d_{insdel}({\bf C})$ of a code ${\bf C} \subset {\bf F}_q^n$ is the minimum of the insdel distances of all different two codewords in this code. Hence we have the direct upper bound $d_{insdel}({\bf C}) \leq 2d({\bf C})$, and the direct Singleton bound on the insdel distance of a linear $[n,k]_q$ code $$d_{insdel} \leq 2(n-k+1),$$   see \cite{Cheng2021,Chen2021}. For linear insertion-deletion codes, the half-Singleton bound $$d_{insdel} \leq 2(n-2k+2),$$ was proposed and proved in \cite{Cheng2021}, also see \cite{Chen2021}. This half-Singleton bound was improved in \cite{JZCW} to $$d_{insdel} \leq 2(n-2k+1), $$ for linear insertion-deletion codes without the all $1$ codeword. In this section, we prove that for both linear and nonlinear insertion-deletion codes, the above upper bounds are far from the tightness when the length are large compared with $q$. This is a direct application of the above Singleton type upper bounds since $d_{insdel}({\bf C}) \leq 2d({\bf C})$.\\

{\bf Corollary 6.1.} {\em If  the length of  a general insertion-deletion code ${\bf C}$ is bigger than the length $n_i$ in Corollary 4.1, the generalized Singleton upper bound II, III, V in Section 4, and the minimum Hamming distance of ${\bf C}$ satisfies the requirements in Corollary 4.1, the generalized Singleton upper bound II, III, V in Section 4, then the size of this insertion-deletion code has to satisfy $$|{\bf C}| \leq q^{n-\frac{m}{4}d_{insdel}}$$ or $$|{\bf C}| \leq q^{n-\frac{t}{4}d_{insdel}}$$.}\\

When $m$ or $t$ is bigger than $6$, this upper bounds are stronger than the direct Singleton bound for general insertion-deletion codes if the lengths are large. \\

\section{Lengths of optimal locally recoverable codes}

For a linear code ${\bf C}$ over ${\bf F}_q$ with length $n$, dimension $k$ and minimum Hamming distance $d$, we define the locality as follows. Given $a \in {\bf F}_q$, set ${\bf C}(i,a)=\{{\bf x} \in {\bf C}: x_i=a\}$, where $i \in \{1,\ldots,n\}$ is an arbitrary coordinate position. The set ${\bf C}_A(i,a)$ is the restriction of ${\bf C}(i,a)$ to the coordinate positions in $A \subset \{1,\ldots,n\}$. The linear code ${\bf C}$ is a locally recoverable code with locality $r$, if for each $i \in \{1,\ldots,n\}$, there exists a subset $A_i \in \{1,\ldots,n\}-\{i\}$ of cardinality at most $r$ such that ${\bf C}_{A_i}(i,a) \cap {\bf C}_{A_i}(i,a')=\emptyset$ for any given $a\neq a'$. A Singleton-like bound for LRC codes $$d \leq n-k+2-\lceil\frac{k}{r}\rceil $$ was proved in \cite{GHSY2011}. It is clear that this upper bound is just the Singleton bound $d \leq n-k+1$ for linear codes when $r\geq k$.  A linear code attaining this upper bound is called a Singleton-optimal locally recoverable code, see \cite{LWXY}. The maximal length of a Singleton-optimal LRC code was studied in \cite{CFXHF,LWXY}. In this section we show that our generalized Singleton upper bounds for Hamming error-correcting codes give a natural upper bound on the lengths of Singleton-optimal LRC codes with a given minimum distance $2R+1$ or $2R+2$.\\

In \cite{CFXHF,LWXY} and references therein, the maximum lengths of Singleton-optimal LRC codes were upper bounded for the given minimum distances $d=5, 6, 7$ with some additional conditions on the locality $r$. For example in \cite{LWXY}, it was proved that for a Singleton-optimal LRC code with the minimum Hamming distance $d=7$, if the locality satisfies $r\geq 6$ and $r=o(n)$, then $n=O(q^{\frac{4r-2}{r+1}})$. In earlier results of \cite{CFXHF} about the length of Singleton-optimal LRC codes of the minimum Hamming distance $5$ and $6$, the strong condition that the recovery sets are disjoint is needed. Our following upper bound is valid for any given minimum Hamming distance $d=2R+1$ or $d=2R+2$, with an additional condition $k \leq rR$. \\

{\bf Corollary 7.1.} {\em For a Singleton-optimal locally recovaerable code ${\bf C}$ of the minimum Hamming distance $2R+1$ or $2R+2$ and the dimension $k\leq Rr$, the length $n$ of ${\bf C}$ has to satisfy $$n \leq cq^{\frac{(2R+1)}{R}}\cdot ln q^{1/R},$$ where $c$ is a constant independent of $q$ and only depending on $R$. }\\

{\bf Proof.} This is direct from the generalized Singleton type upper bound Corollary 4.2.\\

Similarly from generalized Singleton type upper bound VI, we have the following Corollary 7.2.\\

{\bf Corollary 7.2.} {\em For a Singleton-optimal locally recovaerable code ${\bf C}$ of the minimum Hamming distance $2R+1$ or $2R+2$ and the dimension $k\leq Rr$, the length $n$ of ${\bf C}$ has to satisfy $$n \leq R \cdot \frac{q^4-1}{q-1}.$$}\\

A linear $[n, k, d]_q$ code ${\bf C}$ is called $(r, \delta)$ locally recoverable  if for each coordinate position $j \in \{1, \ldots, n\}$ there is a subset $S_j \subset \{1, \ldots, n\}$ satisfying, $j \in S_j$, $|S_j| \leq r+\delta-1$, and the minimum Hamming distance of the punctured code ${\bf C}|_{S_j}$  is at least $\delta$. The above locally recoverable code is just the case of $\delta=2$. The Singleton-like bound for an $(r, \delta)$ locally recoverable code is $$d \leq n-k+1-(\lceil \frac{k}{r} \rceil-1)(\delta-1).$$ The code attaining this bound is called an optimal $(r, \delta)$ locally recoverable code. The maximal lengths of optimal $(r, \delta)$ locally recoverable code was studied in \cite{CMST} and an upper bound was given. We also give an upper bound on the lengths of small dimension $(r, \delta)$ locally recoverable codes using our generalized Singleton type upper bound.\\

{\bf Corollary 7.3.} {\em For an optimal $(r, \delta)$ locally recovaerable code ${\bf C}$ of the minimum Hamming distance $2R+1$ or $2R+2$ and the dimension $k\leq (\frac{R+1}{\delta+1}+1)r$, the length $n$ of ${\bf C}$ has to satisfy $$n \leq cq^{\frac{(2R+1)}{R}}\cdot ln q^{1/R},$$ where $c$ is a constant independent of $q$ and only depending on $R$. Or we have $$n \leq \frac{R(q^4-1)}{q-1}$$ from the generalized Singleton type upper bound VI. }\\

{\bf Proof.} This upper bound follows from the generalized Singleton type upper bound Corollary 4.2 and the generalized Singleton upper bound V.\\

This upper bound on the lengths of optimal $(r, \delta)$ locally recoverable codes is stronger than the general upper bound in \cite{CMST} when the dimensions are small.\\

\section{Conclusion}

In this paper, we proposed that any small covering code ${\bf C}'$ in a general finite metric space with the covering radius $R({\bf C}') \leq d_{list}$ gives a good upper bound on the sizes of $(d, L)$ list-decodable codes. When the list size is $L=1$, many Singleton type upper bounds on the sizes of codes with a given minimum Hamming distance $d$ can be obtained from small covering codes with the covering radius $\leq \frac{d-1}{2}$. These upper bounds are stronger than the Griesmer bound and some previous upper bounds when the lengths are large. From our generalized Singleton type upper bounds, some upper bounds on the length of general small Singleton defect codes were presented. We also showed that many previous constructions of linear codes with few nonzero weights give natural covering codes from the Delsarte bound on covering radii. As applications, generalized Singleton type upper bounds for long insertion-deletion codes were given. Upper bounds on the lengths of small dimension optimal locally recoverable codes and small dimension $(r, \delta)$ locally recoverable codes with any given minimum distance were also presented.\\

\end{document}